\documentclass[conference]{IEEEtran}
\usepackage{hyperref}
\usepackage[cp1250]{inputenc}
\usepackage{amsmath}
\usepackage{amssymb}
\usepackage{graphicx}
\usepackage{url}
\usepackage{booktabs}
\usepackage{array}
\title{Frequency Effects on Predictability of Stock Returns}
\author{\IEEEauthorblockN{Paweł Fiedor}
\IEEEoverridecommandlockouts
\thanks{This pre-print has been submitted to the IEEE for possible publication.}
\IEEEauthorblockA{Cracow University of Economics\\
Rakowicka 27, 31-510 Kraków, Poland\\
Email: s801dok@wizard.uek.krakow.pl}
}
\begin{document}
\maketitle
\begin{abstract}
We propose that predictability is a prerequisite for profitability on financial markets. We look at ways to measure predictability of price changes using information theoretic approach and employ them on all historical data available for NYSE 100 stocks. This allows us to determine whether frequency of sampling price changes affects the predictability of those. We also relations between price changes predictability and the deviation of the price formation processes from iid as well as the stock's sector. We also briefly comment on the complicated relationship between predictability of price changes and the profitability of algorithmic trading.
\end{abstract}

\section{Predictability in social sciences}

Prices are the main object of research in finance. Even though since the 1980s the support for the strong version of the efficient market hypothesis has dwindled \cite{Lee:2009,Yen:2008}, price formation and price observation are the main bases for all research in finance, and a large part of economics.

Researchers in social sciences (much like their counterparts in natural sciences) are often interested in crucial questions about predictability, which (also in economics and finance) are not trivial due to the human involvement \cite{Rosser:2008}. Compared to social systems the ones studied by physics are often relatively trivial (in many cases being the subset of the socioeconomic systems), nonetheless the methods of physics (particularly complex systems physics) are often helpful to understand the complexity of social systems, which led to the emergence of the field of econophysics in the last 20 years \cite{Mantegna:2000}.

While researchers in humanities are interested in predictability of many processes, such as movements of people and their communication patterns \cite{Song:2010} the economists and financial researchers are most interested in the predictability of prices and their changes, due to the importance of prices in economics and finance and due to the vast amounts of data which can be analysed in this effort.

Traditionally econometricians and econophysicists are more interested in predictability of price changes in principle and in practice \cite{Navet:2008}. Financial engineers on the other hand are more interested in profitability, regardless of predictability \cite{Youwei:2005,Kohli:2011}. Nonetheless we propose that question of predictability is more fundamental and is crucial to the problem of profitability of trading algorithms (which would be the derivative of the characteristics of price formation processes), whether it is enhancing or hindering it.

There are some efforts to bridge the gap between profitability and predictability, none of which have yielded unassailable results \cite{Wu:2004,Rothenstein:2004,Navet:2008}. Heuristically it seems that predictability deals with the existence of patterns in price formation processes, which despite being more fundamental than profitability, does not necessarily help it. It is conceivable that certain trading strategies will not be able to find the patterns even if they exist, therefore the predictability of price formation processes will be oblivious to profitability of those strategies or algorithms. But if the price formation processes are highly entropic then there are no patterns to be found and all trading strategies will rest on randomness, arbitrage or information asymmetries. Some investors or trading strategies may indeed prefer less or more entropic price formation processes however, so the predictability of stocks may be highly relevant to the choice of trading strategy indeed.

Though particularly with high-frequency trading the question is, assuming time series is predictable, whether it will not be more profitable for the agents with the best access to the trading platform (lowest time between input and output) \cite{Menkveld:2011}. Therefore we are looking at price history at different time intervals, that is standard daily price changes and intraday price changes capturing all price changes recorded at the stock exchange, regardless of their frequency.

\section{Methodology}

Predictability can be measured using the concept of entropy. Low entropy indicates high certainty and predictability, whereas high entropy indicates low predictability and information availability.

\subsection{Entropy}

The predictability of a time series can be estimated using the entropy rate. Entropy rate is a term derivative to the notion of entropy, which measures the amount of uncertainty in a random variable. The Shannon's entropy of a single random variable $X$ is defined as 
\begin{equation}
	\label{eq:Def_entropy}
H(X) = -\sum_{i} p(x_i) \log_2 p(x_i) 
\end{equation}
summed over all possible outcomes $\{x_i\}$ with respective probabilities of $p(x_i)$ \cite{Shannon:1948}. 
For two random variables $(X,Y)$, \emph{joint entropy} 
$H(X,Y)$ measuring the uncertainty associated with both, and \emph{conditional entropy} $H(X|Y)$ measuring uncertainty in one random variable assuming the other has been observed, can be calculated.  The joint entropy and conditional entropy are related in a following manner: 
\begin{equation}
	\label{eq:Ent_joint_cond}
H(X|Y) = H(X,Y) - H(Y)
\end{equation}

Shannon also introduced the \emph{entropy rate}, which generalises the notion of entropy for sequences of \emph{dependent} random variables. 
For a stationary stochastic process $X = \{X_i\}$, the entropy rate is 
defined as
\begin{equation}
	\label{eq:Def_entropy_rate}
	H(X) = \lim_{n \rightarrow \infty} \frac{1}{n} H(X_1, X_2, \dots, X_n)
\end{equation}
\begin{equation}
	\label{eq:Def_entropy_rate_2}
	H(X) = \lim_{n \rightarrow \infty} H(X_n | X_1, X_2, \dots, X_{n-1})
\end{equation}
where \eqref{eq:Def_entropy_rate} holds true for all stochastic processes, but \eqref{eq:Def_entropy_rate_2} requires stationarity of the process. 
The right side of \eqref{eq:Def_entropy_rate} can be interpreted such as that entropy rate measures the uncertainty in a quantity at 
time $n$ having observed the complete history up to that point. Theory of information defines entropy rate of a stochastic process as the amount of new information created in a unit of time \cite{Cover:1991}. Hence entropy rate can be interpreted as maximum rate of information creation which can be processed as price changes. Nonetheless the entropy rate denotes the average entropy of each random variable in the stochastic process and in this study we use this weaker interpretation of the entropy rate to characterise the average uncertainty of a quantity at any given time. Joint and conditional entropy rates can similarly be defined and interpreted.

\subsection{Entropy estimation}

Estimation of the entropy (rate) is greatly relevant due to the fact that the real entropy is known in very few isolated applications. Research in entropy estimation has been quite active in the past 15 years, especially due to the advances in neurobiology and the usefulness of entropy together with similar information-theoretic constructs to studying the activities of the brain (most notably EEG \cite{Maciejewski:2008}).
Methods of entropy estimation can be grouped into two separate categories \cite{Gao:2006}:
\begin{enumerate}
\item Maximum likelihood estimators (plug-in estimators), which study empirical distribution of all words of a given length (usually constructing Markov chain or order $n$ and calculating its entropy). This approach has a drawback in the exponentially increasing requirements for the sample size for higher word lengths. This means that those estimators are not practical for analysing the mid- and long-term relations, which cannot be ignored in economics and finance. Therefore those methods are becoming less popular and we are not using them in this study.
\item Estimators based on data compression algorithms, most notably Lempel-Ziv algorithm \cite{Farah:1995,Kontoyiannis:1998a,Lempel:1977} and Context Tree Weighting \cite{Willems:1995,Kennel:2005}. Both methods are precise even for a limited sample size \cite{Louchard:1997,Leonardi:2010} (which we demonstrate later), therefore are better equipped to deal with mid- and long-term relationships in the analysed data. In this study we use both those methods and compare them, providing evidence for their relative usefulness, which is not yet fully explained in the literature despite some reporting on slightly better characteristics of CTW \cite{Navet:2008,Gao:2008}.
\end{enumerate}

\subsection{Lempel-Ziv complexity}

Complexity in the Kolmogorov sense can be used to estimate entropy rate. In 1965 Kolmogorov defined the complexity of a sequence as the size of the smallest binary program which can produce this sequence \cite{Cover:1991}. This definition is not operational, therefore intermediate measurements are used, which are designed to measure complexity in the Kolmogorov sense. Lempel-Ziv algorithm is one of those measurements, which test the randomness of series. This algorithm measures linear complexity and has been first introduced by Jacob Ziv and Abraham Lempel in 1977 \cite{Lempel:1977}. This measurement counts the number of patterns in the series, scanning it from left to right, so for example the complexity the below series:
\[s = 101001010010111110\]
is equal to 8, since scanning it from left to right one finds 8 distinct patterns \cite{Doganaksoy:2006}:
\[1|0|10|01|010|0101|11|110|\]
On this basis there have been a number of estimators of entropy rate created. In this article we follow \cite{Navet:2008} and use the estimator created by Kontoyiannis in 1998 (estimator $a$) \cite{Kontoyiannis:1998}. This estimator is widely used \cite{Kennel:2005,Navet:2008} and it was shown that it has better statistical properties than previous estimators based on Lempel-Ziv algorithm \cite{Kontoyiannis:1998}, though there is a large choice of slightly different variants to choose from \cite{Gao:2008}, which is largely irrelevant.

More formally, to calculate the entropy of a random variable $X$, the probability of each possible outcome $p(x_i)$ must be known. When these probabilities are not known, entropy can be estimated by replacing the probabilities with relative frequencies from observed data. Estimating the entropy rate of a stochastic process is more complex as random variables in stochastic processes are usually interdependent. Then the mentioned estimator is defined as:

\begin{equation}
	\label{eq:LZ_complexity}
	\hat{H_{lz}} = \frac{n \log_2 n}{\sum_i \Lambda_i},
\end{equation}
where $n$ denotes the length of the time series, and $\Lambda_i$ denotes the 
length of the shortest substring starting from time $i$ 
that has not yet been observed prior to time $i$, i.e. from time $1$ to 
$i-1$. 
It is known that for stationary ergodic processes, $\hat{H}(X)$ converges to 
the entropy rate $H(X)$ almost surely as $n$ approaches infinity \cite{Kontoyiannis:1998}. For the purpose of this study we have implemented this estimator using C++.

\subsection{Context Tree Weighting}

For any discrete-time stationary and ergodic stochastic process $X$ asymptotic equipartition property (proven for finite-valued stationary ergodic sources in Shannon-McMillan-Breiman theorem) asserts that:
\begin{equation}
	\label{eq:AEP}
    -\frac{1}{n} \log p(X_1^n) \to H(X) \quad \mbox{ as } \quad n\to\infty
\end{equation}
where $p(X_1^n)$ denotes the probability of process $X_1^n$ limited to duration $\{1,\dots, n\}$, and $H(X)$ is the entropy rate of $X$, and is shown to exist for all discrete-time stationary processes. The convergence is proven with probability of $1$ in all cases \cite{Cover:1991}. Therefore we can estimate $H$ through estimating probability of a long realisation of $X$.

Context Tree Weighting (CTW) is a data compression algorithm \cite{Willems:1995,Willems:1996,Willems:1998}, which can be interpreted as a Bayesian procedure for estimating the probability of a string generated by a binary tree process \cite{Gao:2008}.

A binary tree process of depth $D$ is a binary stochastic process $X$ with a distribution defined with a suffix set $S$ consisting of binary strings of length $ \leq D$ and a parameter vector $\Theta = (\Theta_s ; s \in S)$, where each $\Theta_s \in [0;1]$.

If a certain string $x_1^n$ has been generated by a tree process of depth $\leq D$, but with unknown suffix set $S^*$ and parameter vector $\Theta^*$ then we may assign a prior probability $\pi(S)$ on each suffix set $S$ of depth
$\leq D$ and, given $S$, we may assign a prior probability $\pi(\Theta | S)$ on each parameter vector $\Theta$. A Bayesian approximation to the true probability of $x_1^n$ (under $S^*$ and $\Theta^*$) is the mixture probability:

\begin{equation}
	\label{eq:Mixture_prob}
\hat{P}_{D, mix}(x_1^n)=\sum_s{\pi(S)}\int{P_{S,\Theta}(x_1^n)\pi(\Theta | S) d\Theta}
\end{equation}

where $P_{S,\Theta}(x_1^n)$ is the probability of $(x_1^n)$ under the distribution of a tree process with suffix set $S$ and parameter vector $\Theta$. The expression in \eqref{eq:Mixture_prob} is impossible to compute directly, since the number of suffix sets of depth $\leq D$ is of order $2^D$. This is prohibitively large for practical use for any $D > 20$.

The CTW algorithm is an efficient way of computing the mixture probability in \eqref{eq:Mixture_prob} given a specific choice of the prior distributions $\pi(S), \pi(\Theta | S)$. The prior on $S$ is
\begin{equation}
	\label{eq:Prior}
\pi(S)=2^{-|S|-N(S)+1}
\end{equation}
where $|S|$ is the number of elements of $S$ and $N(S)$ is the number of strings in $S$ with length strictly smaller than $D$. Given a suffix set $S$, the prior on $\Theta$ is the product $(\frac{1}{2},\frac{1}{2})$-Dirichlet distribution, i.e., under $\pi(\Theta | S)$ the individual $\Theta_s$ are independent, with each $\Theta_s \sim$ Dirichlet$(\frac{1}{2},\frac{1}{2})$.

It is of paramount importance that the CTW algorithm is able to compute the probability in \eqref{eq:Mixture_prob} precisely. This computation can be performed in time and amount of memory used growing linearly with the length of the string $n$. Therefore it is possible to study lengths of $D$ much higher than it is possible with the maximum likelihood estimators.

Therefore given a binary string $x_1^n$ CTW entropy rate estimator $\hat{H_{ctw}}$ is given:
\begin{equation}
	\label{eq:Entropy_CTW}
\hat{H_{ctw}}=-\frac{1}{n}\log{\hat{P}_{D,mix}(x_1^n)}
\end{equation}
where $\hat{P}_{D,mix}(x_1^n)$ is the mixture probability in \eqref{eq:Mixture_prob} computed by the CTW algorithm \cite{London:2002,Kennel:2002}. For the purpose of this study we have modified the algorithm developed by Frans Willems, Yuri Shtarkov and Tjalling Tjalkens in cooperation with KPN NV (available at \url{http://www.ele.tue.nl/ctw/}), in C.

\section{Algorithm testing}

Later in this paper we will compare how two presented methods work applied to financial data, but before that we look at their characteristics on data for which we know the real entropy in advance (at least with good approximation). That way we can see how fast those methods converge to the real entropy with sample increasing in length. First we look at fully predictable set of data, which is a series of zeroes of a given length (entropy equal to $0$). Then we look at a more meaningful example, that is a fully random sequences being realisations of a random variable $P$ with uniform distribution among $\{0,1,2,3\}$. Theoretical entropy for such variable is given by: \[H(P)=-\sum_{i=0}^{3}\frac{1}{4}log_2\frac{1}{4}=2\] A result would then be dependent on the sample size, the efficiency of the estimator (two described previously) as well as the quality of the random generator (here we assume the quality to be perfect within measurement errors). As the generator we use random integer generator based on atmospheric noise available at \url{http://www.random.org}. This generator has been successfully used in a number of research projects \cite{Biggar:2008,Zijlstra:2013}. We average those results over 10 realisations of $P$.

We may particularly be interested in how fast the entropy rate estimation converges to the real value with sample size, as those are usually limited, especially when analysing daily price changes.

\begin{figure}[tbh]
\centering
\includegraphics[width=0.5\textwidth]{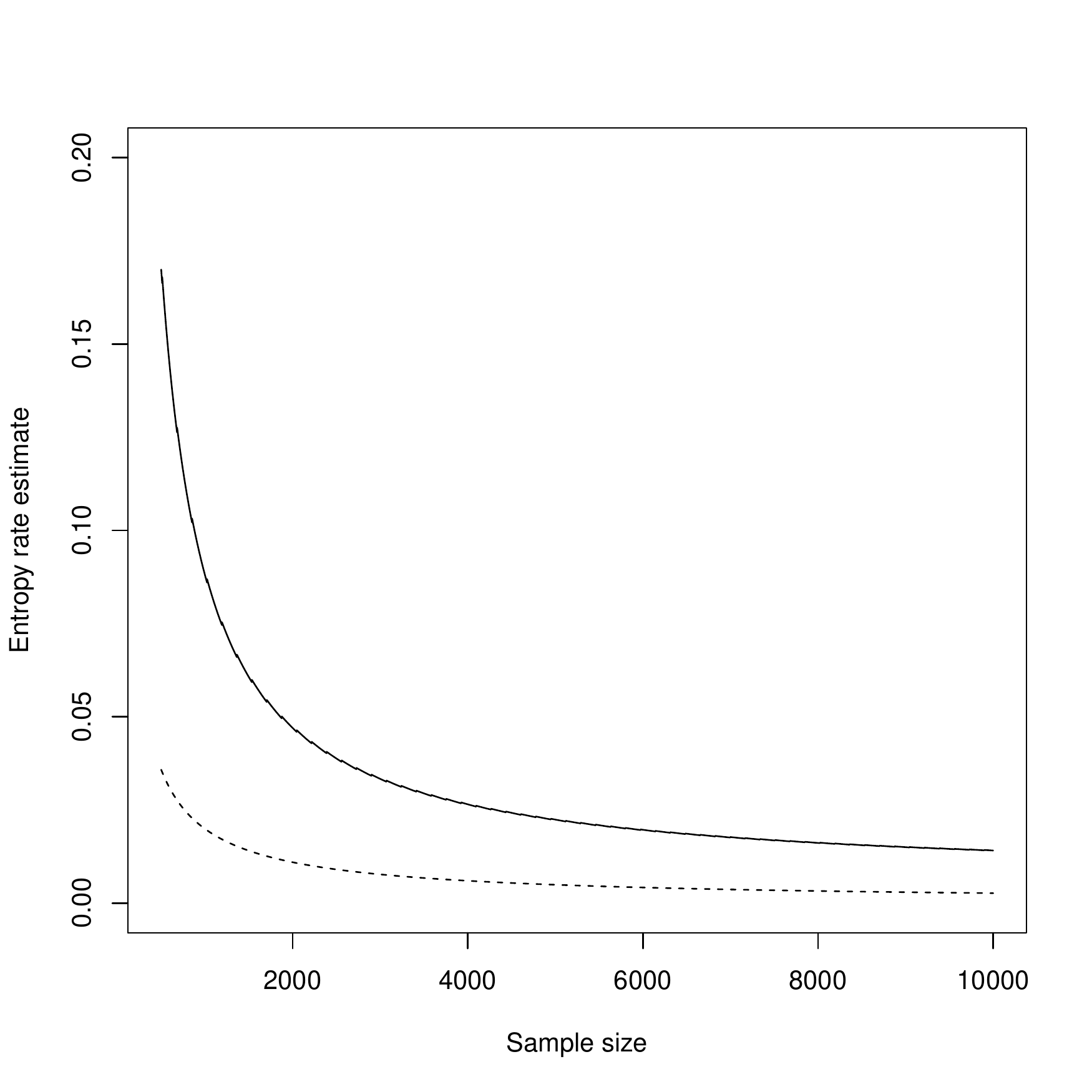}
\caption{Algorithm comparison for non-random sample}
\label{fig:TestNRand}
\end{figure}

The results for process with a priori known entropy rate of 0 are shown on Figure \ref{fig:TestNRand}. As can be seen Lempel-Ziv (dashed line) algorithm converges to the real value more quickly and is closer to it at all times than Context Tree Weighting. The latter seems to be overestimating entropy rate for low entropic processes.

\begin{figure}[tbh]
\centering
\includegraphics[width=0.5\textwidth]{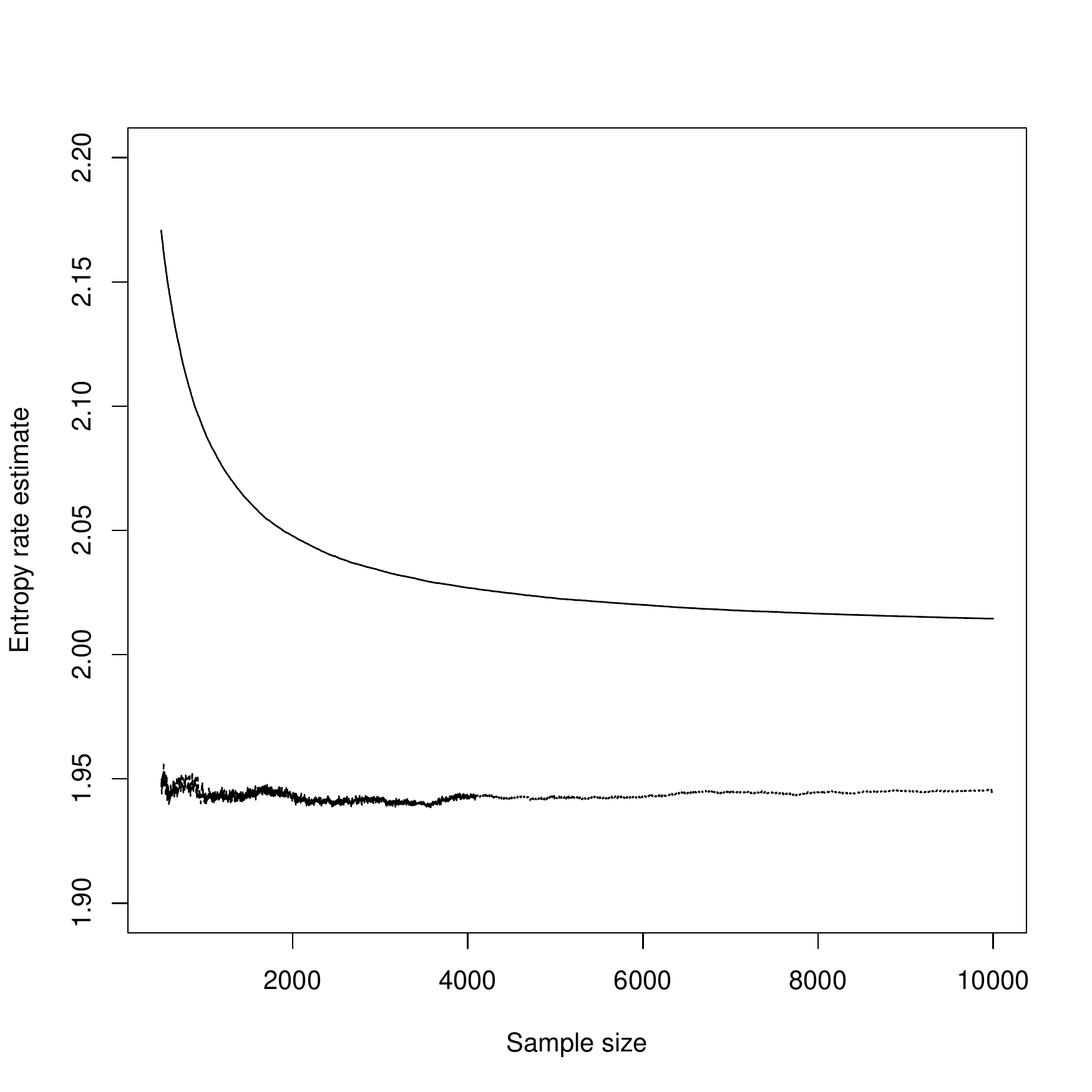}
\caption{Algorithm comparison for random sample}
\label{fig:TestRand}
\end{figure}

The results for random process (a priori entropy rate $\sim{}2$) are shown on Figure \ref{fig:TestRand}. As can be seen Context Tree Weighting algorithm (line above) overestimates the value slightly, which Lempel-Ziv algorithm underestimates the value but to a smaller degree than CTW. It also can be seen that Lempel-Ziv algorithm converges to the real value faster than Context-Tree Weighting. Calculations for samples of size under 4000 with Context Tree Weighting seems unwise, for larger samples the choice appears irrelevant keeping in mind that CTW will give slightly higher results than L-Z.

\section{Data study --- GPW}

In our study we estimate the entropy of the daily price time series of 91 securities traded on New York Stock Exchange (NYSE100) (the 9 missing stocks were excluded due to missing data). The data has been downloaded from Google Finance database available at \url{http://www.google.com/finance/} and was up to date as of the 11th of November 2013, going 10 years back. The data is transformed in the standard way for analysing price movements, that is so that the data points are the log ratios between consecutive daily closing prices: $r_{t}=ln(p_{t}/p_{t-1})$ and those data points are, for the purpose of the estimators, discretized into 4 distinct states. The states represent 4 quartiles, therefore each state is assigned the same number of data points. This design means that the model has no unnecessary parameters, which could affect the results and conclusions reached while using the data. This and similar experimental setups have been used in similar studies \cite{Navet:2008} (Navet \& Chen divided data into 8 equal parts instead of quartiles) and proved to be very efficient at revealing the randomness of the original data, which is the main purpose of the study \cite{Steuer:2001}.

\begin{figure}[tbh]
\centering
\includegraphics[width=0.5\textwidth]{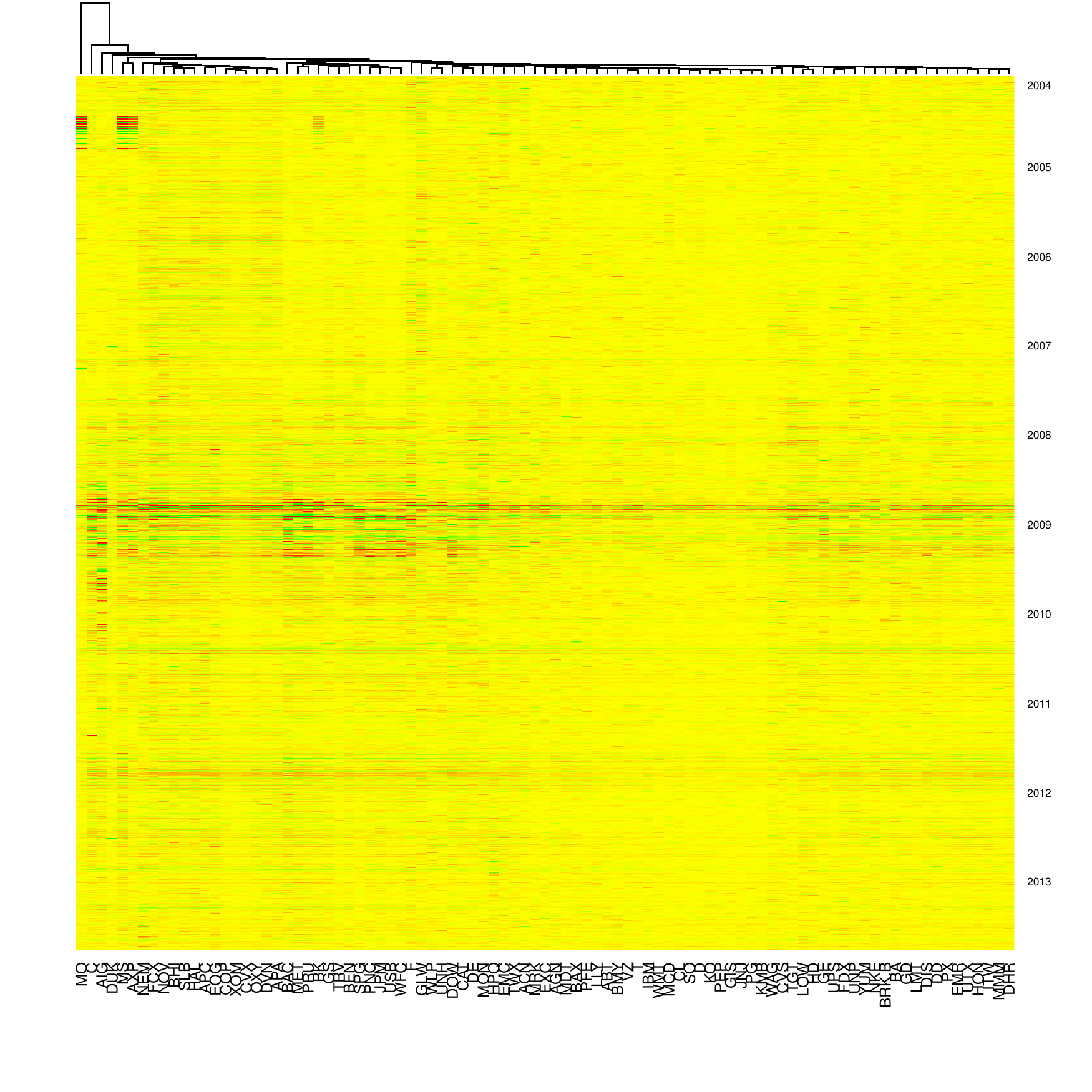}
\caption{{Heatmap for NYSE100 daily log price changes over time}}
\label{fig:heatmap}
\end{figure}

The original log returns are shown on Figure \ref{fig:heatmap}. Green denotes low values, while red high values, with yellow in the middle of the scale. The columns are ordered according to the dendrogram on the top, calculated using hierarchical clustering. It's not noticeable on Figure \ref{fig:heatmap} but the companies are often clustered according to their sectors. 

The main question asked in this paper is whether daily price changes are more or less predictable than intraday (high-frequency) price changes. Heuristically one would imagine that daily price changes are less incidental and should be therefore more predictable, which is the hypothesis. To compare the predictability of daily price changes with the predictability of high frequency price changes we have also estimated entropy for another set of data, that is intraday (1-minute intervals) price changes for the same 91 securities listed on NYSE100. The data covers 15 days between the 21st October 2013 and the 8th of November 2013. Therefore the length of time series in both cases are comparable and sufficient for the used algorithms.

\section{Results}

\begin{figure}[tbh]
\centering
\includegraphics[width=0.5\textwidth]{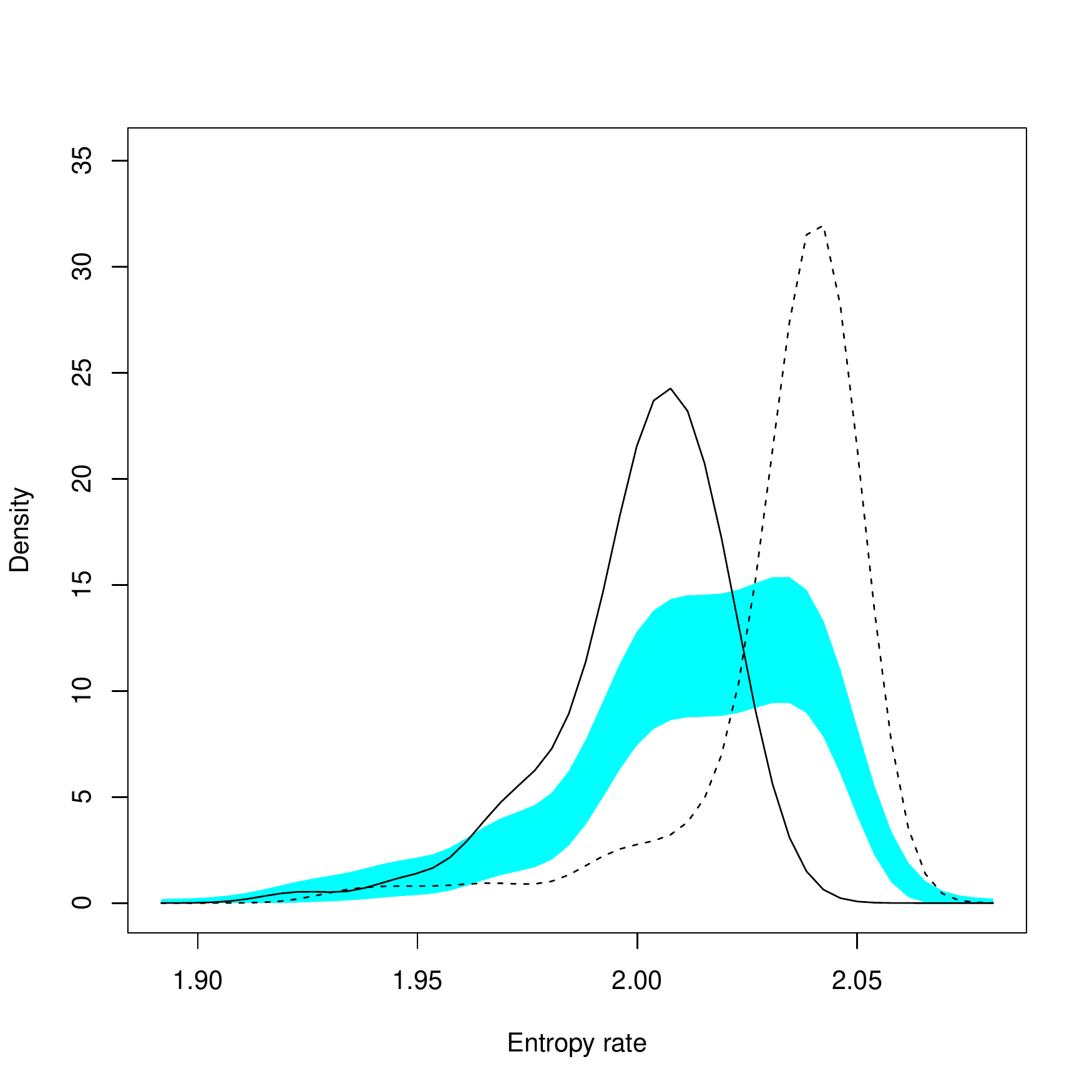}
\caption{Kernel density for daily and intraday entropy rates (CTW)}
\label{fig:KernelCTW}
\end{figure}

\begin{figure}[tbh]
\centering
\includegraphics[width=0.5\textwidth]{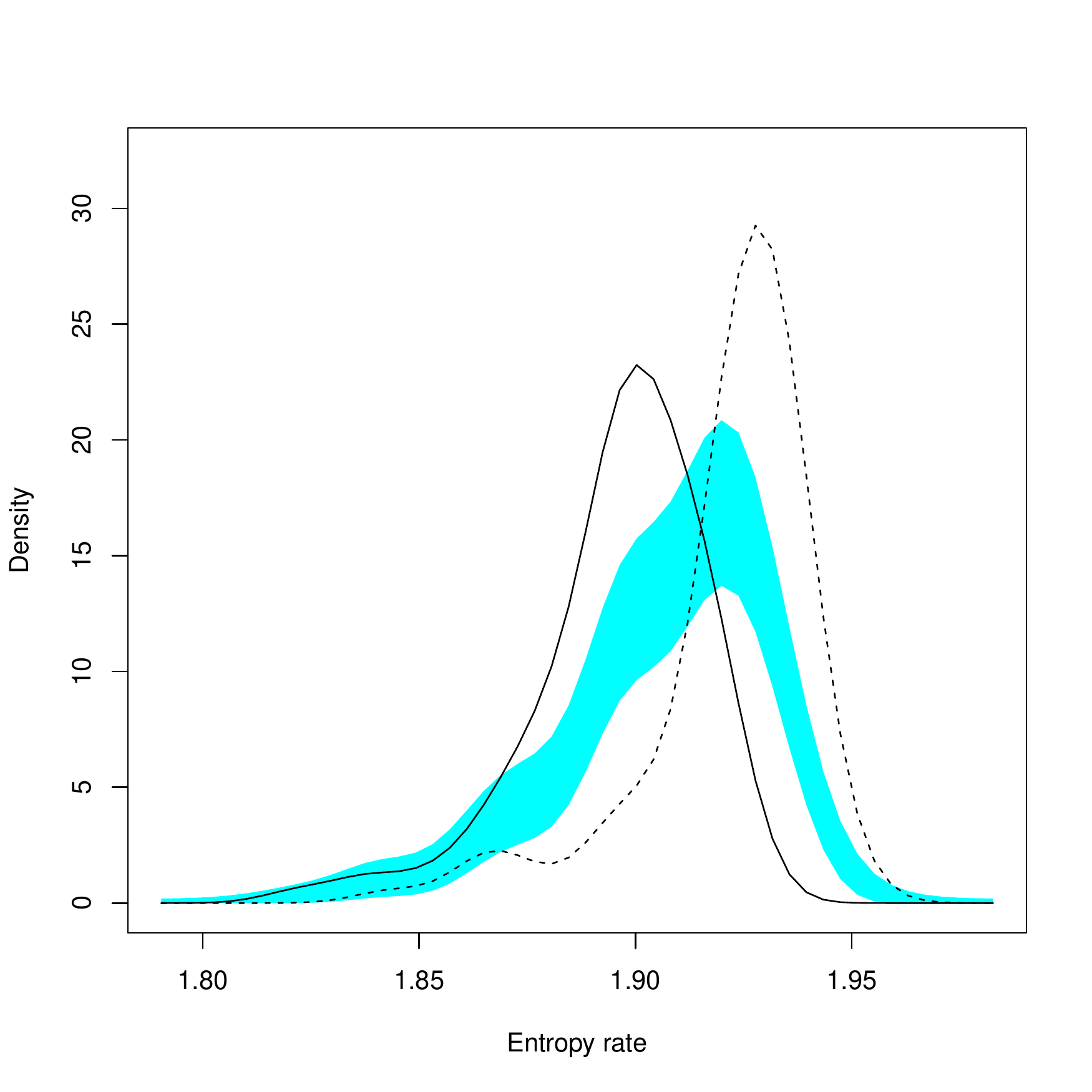}
\caption{Kernel density for daily and intraday entropy rates (L-Z)}
\label{fig:KernelLZ}
\end{figure}

Kernel densities for Context Tree Weighting entropy rate estimates for daily (dashed lines) and intraday (solid lines) together with a reference band used in testing their equality are shown on Figure \ref{fig:KernelCTW}. Similarly those values for entropy rates calculated using Lempel-Ziv algorithm are presented on Figure \ref{fig:KernelLZ}. Our study revealed that the average estimated entropy rate of daily price changes is equal to $2.04$ (CTW) or $1.92$ (Lempel-Ziv) out of the theoretical maximum of $2$, which shows that those price changes are not random, but are not easily predictable (standard deviation of $0.03$ and $0.05$ show that the predictability is robust). Surprisingly the average estimated entropy for intraday price changes is equal to $2$ (CTW) or $1.90$ (Lempel-Ziv), which means that high frequency price changes are more predictable than daily price changes.

We have tested the kernel densities for equality, and have obtained p-values of nearly 0, hence we can refute the hypothesis stating equality in daily and intraday entropy rate distributions among NYSE100 stocks. Graphically it's presented on Figures \ref{fig:KernelCTW} and \ref{fig:KernelLZ} as reference bands. Had those been statistically equal both densities would be presented within the boundaries of the reference bands.

\begin{figure}[tbh]
\centering
\includegraphics[width=0.5\textwidth]{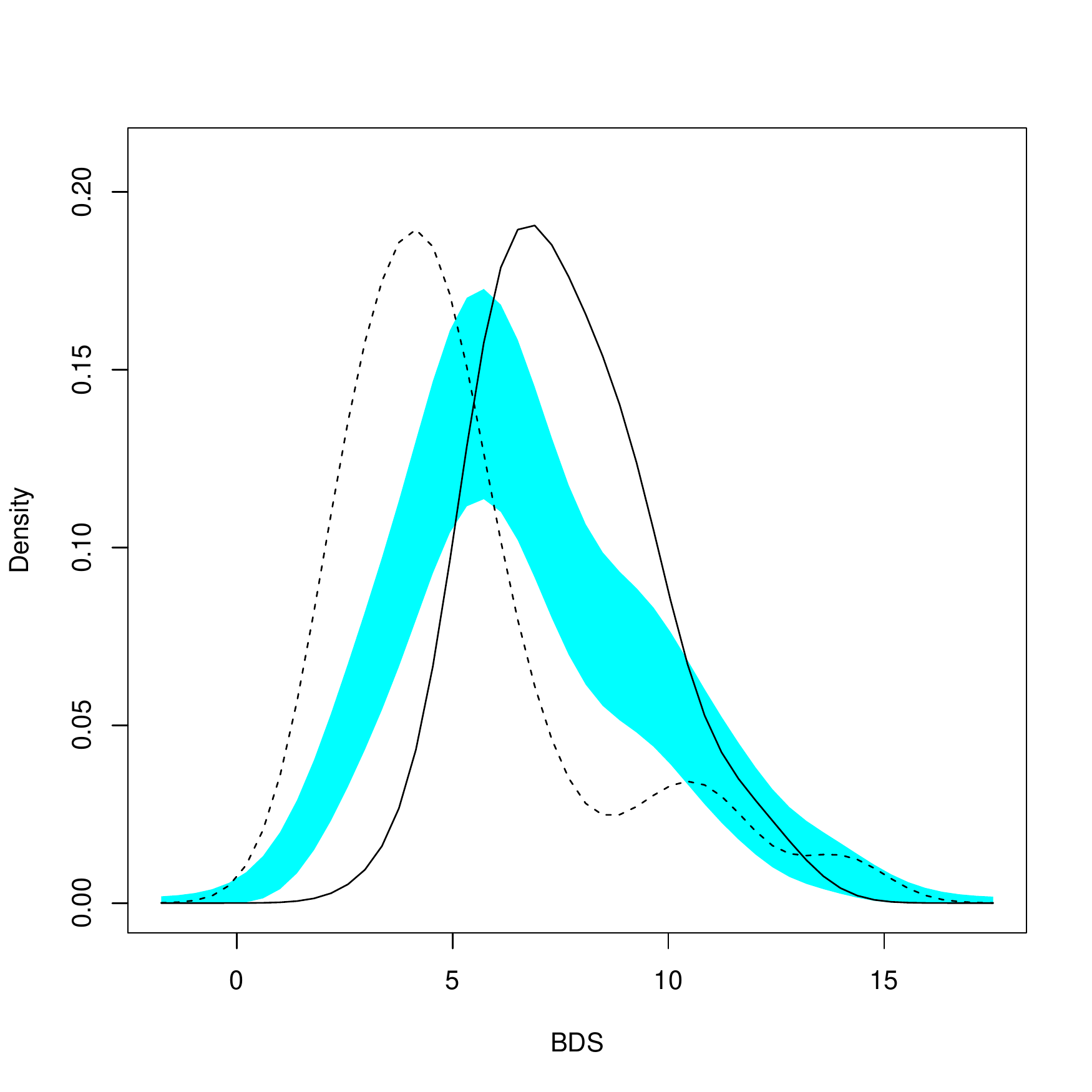}
\caption{Kernel density for daily and intraday BDS}
\label{fig:KernelBDS}
\end{figure}

\begin{figure}[tbh]
\centering
\includegraphics[width=0.47\textwidth]{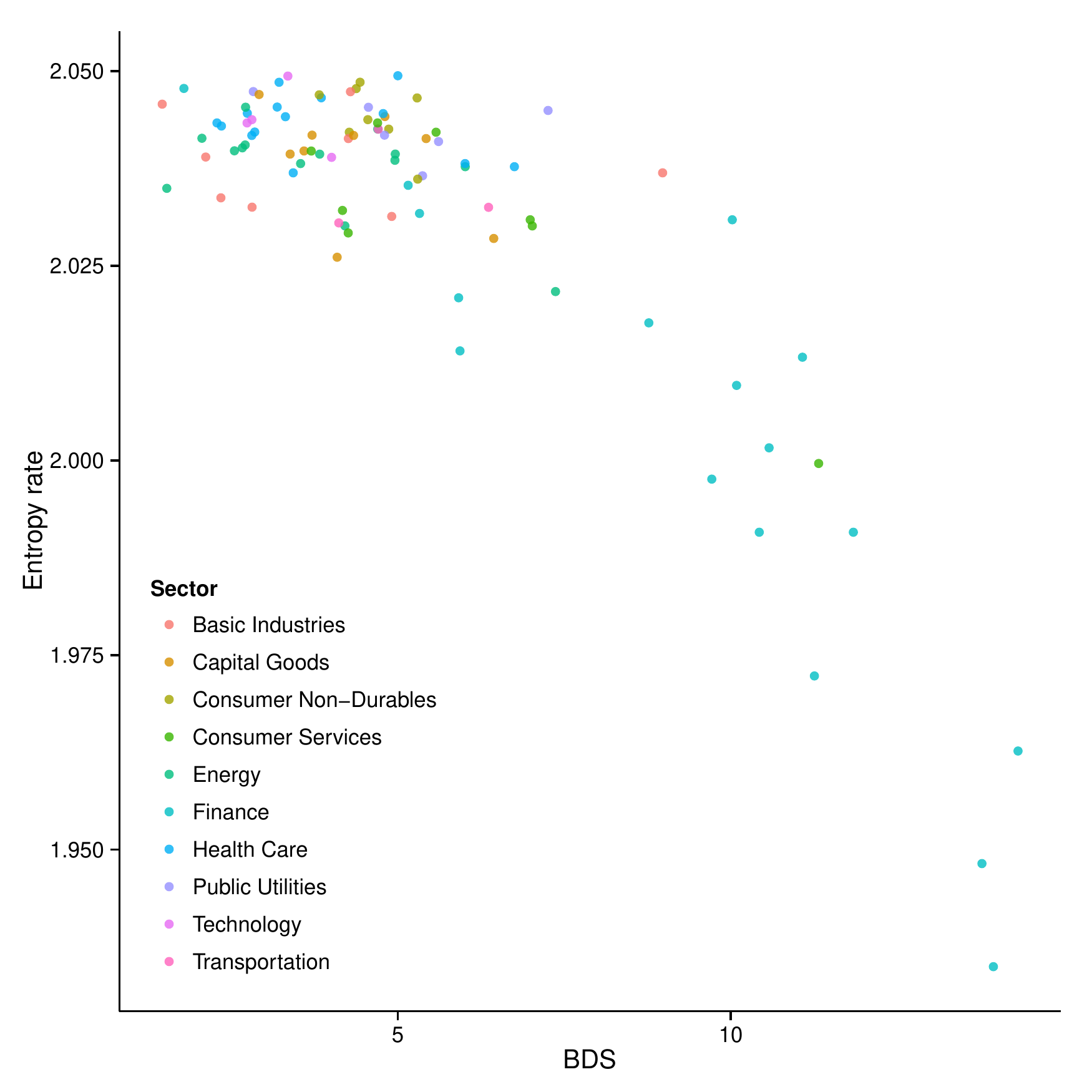}
\caption{Scatterplot for daily entropy rates (CTW) and BDS}
\label{fig:ScatterDailyBDS}
\end{figure}

\begin{figure}[tbh]
\centering
\includegraphics[width=0.47\textwidth]{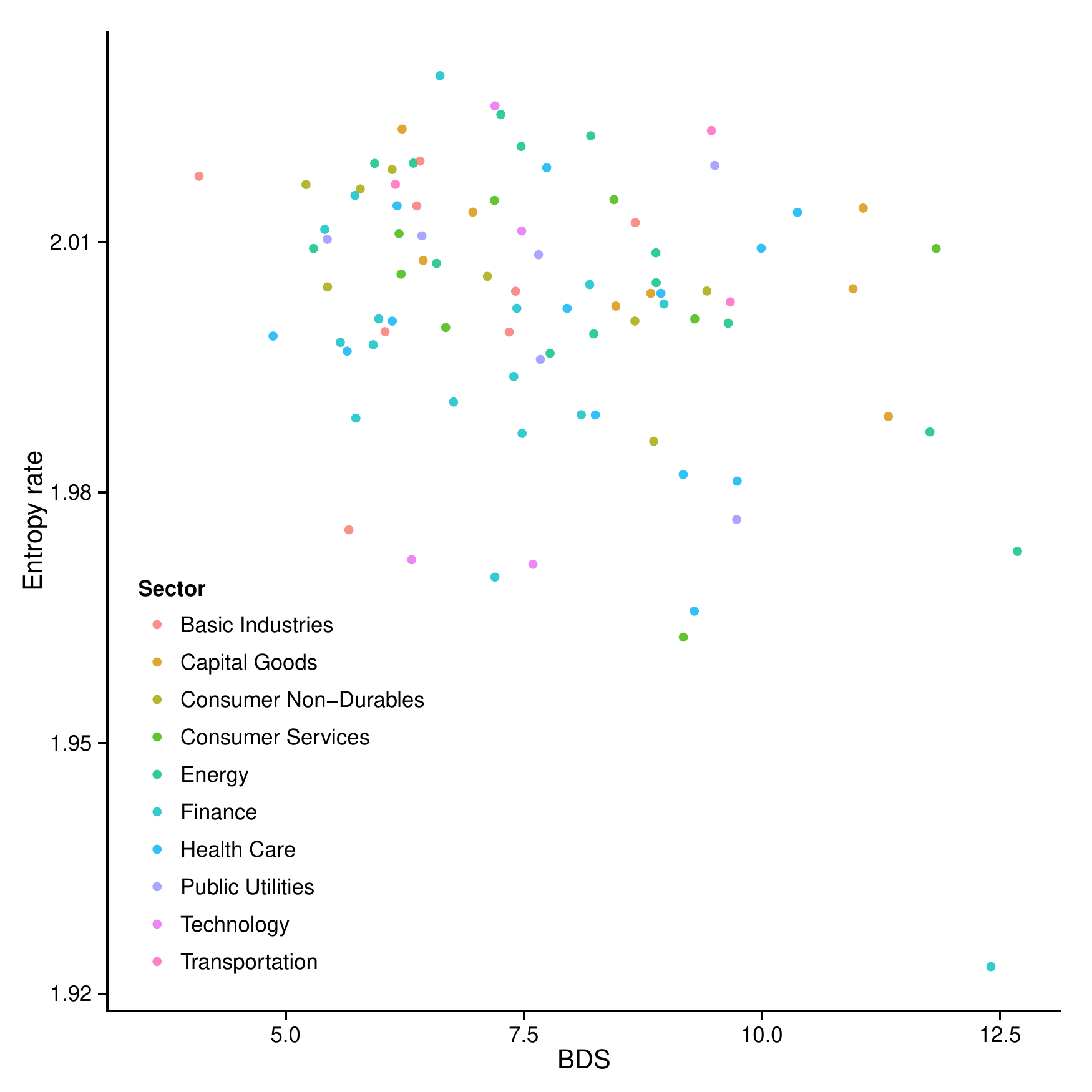}
\caption{{Scatterplot for intraday entropy rates (CTW) and BDS}}
\label{fig:ScatterIDBDS}
\end{figure}

We use BDS statistics \cite{Hsieh:1993,Brock:1995,Chu:2001} to test the hypothesis that daily and intraday log price changes are iid processes. The test takes into consideration both linear and nonlinear dependencies causing the process to deviate from iid property. For the great majority of studied companies the log price changes are not iid, but nonetheless it's more interesting in this analysis to see at how far the deviation goes (therefore the choice of BDS parameters is not of paramount importance as long as it's consistent). BDS statistic takes standard Normal distribution. On Figure \ref{fig:KernelBDS} we can see that intraday log price changes are deviating from iid more than daily ones, which is consistent with them being more predictable. We can see on Figures \ref{fig:ScatterDailyBDS} \& \ref{fig:ScatterIDBDS} how the BDS statistic correlates with the entropy rate for the same company, for daily and intraday price changes respectively. We can see that the less predictable a price formation process is the closer it is to being iid, which is to be expected.

\begin{figure}[tbh]
\centering
\includegraphics[width=0.47\textwidth]{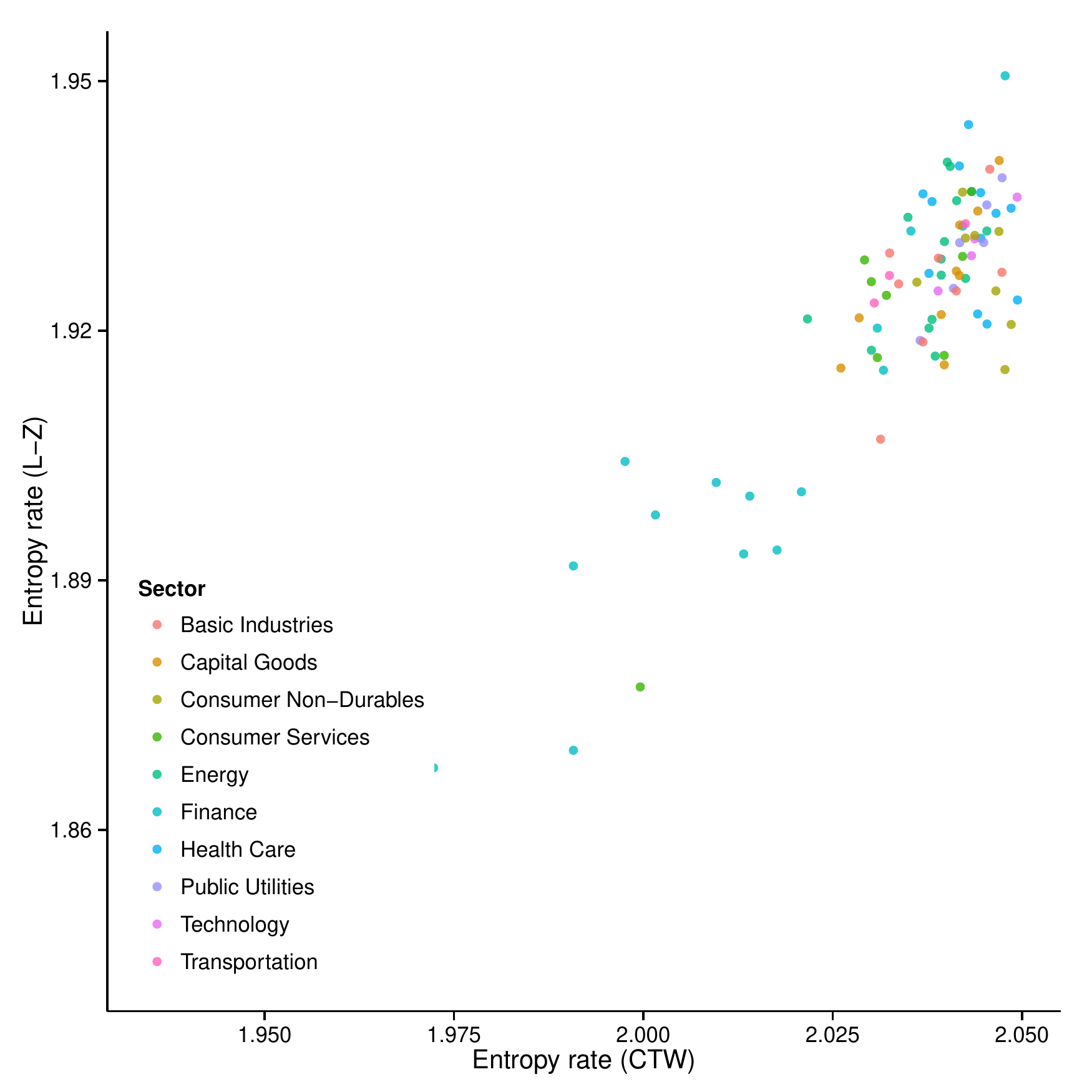}
\caption{Scatterplot for daily entropy rates (CTW and L-Z)}
\label{fig:ScatterDaily}
\end{figure}

\begin{figure}[tbh]
\centering
\includegraphics[width=0.47\textwidth]{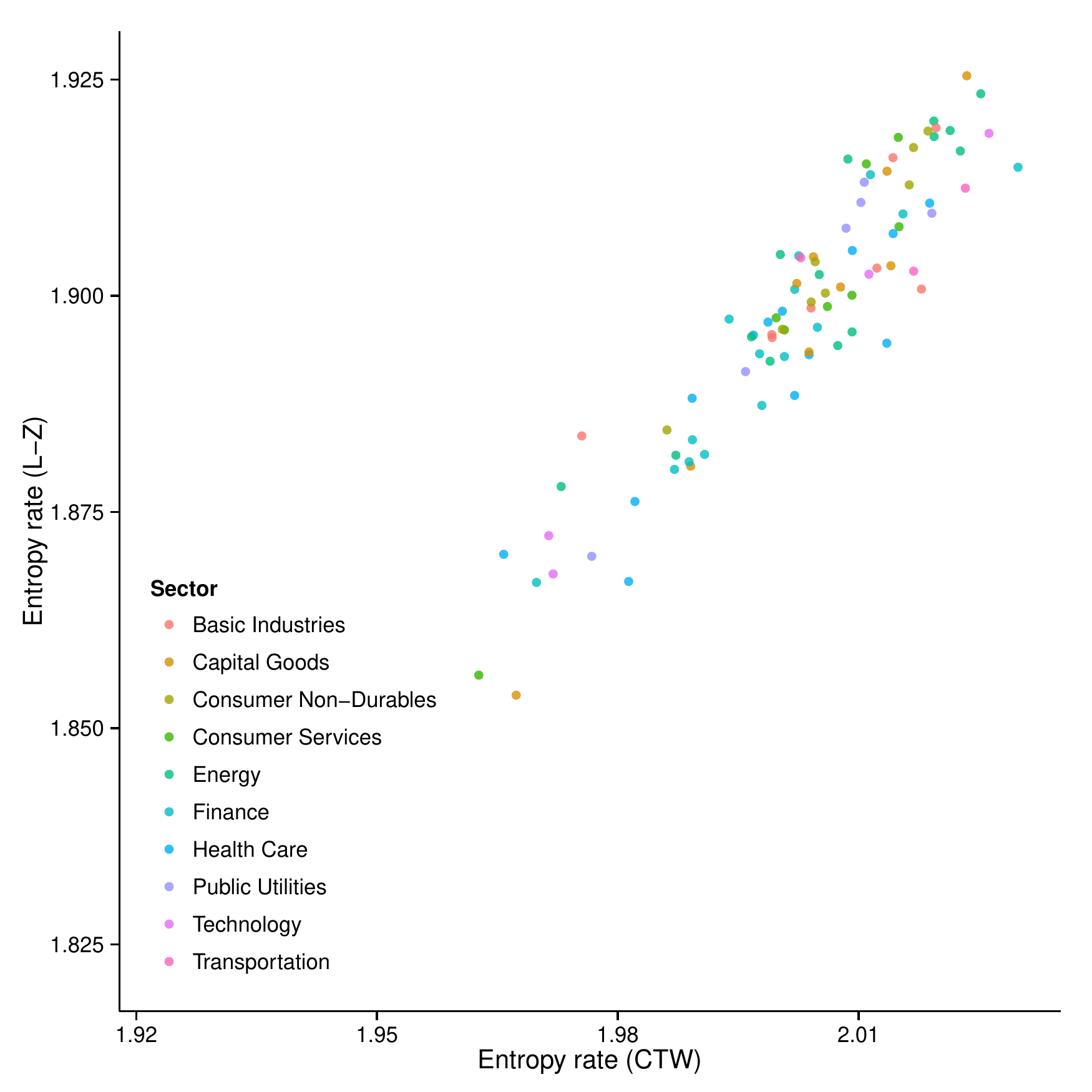}
\caption{{Scatterplot for intraday entropy rates (CTW and L-Z)}}
\label{fig:ScatterID}
\end{figure}

Further we can see on Figures \ref{fig:ScatterDaily} \& \ref{fig:ScatterID} that both used algorithms are indeed highly correlated, proving further their quality in estimating entropy rates.

\begin{figure}[tbh]
\centering
\includegraphics[width=0.5\textwidth]{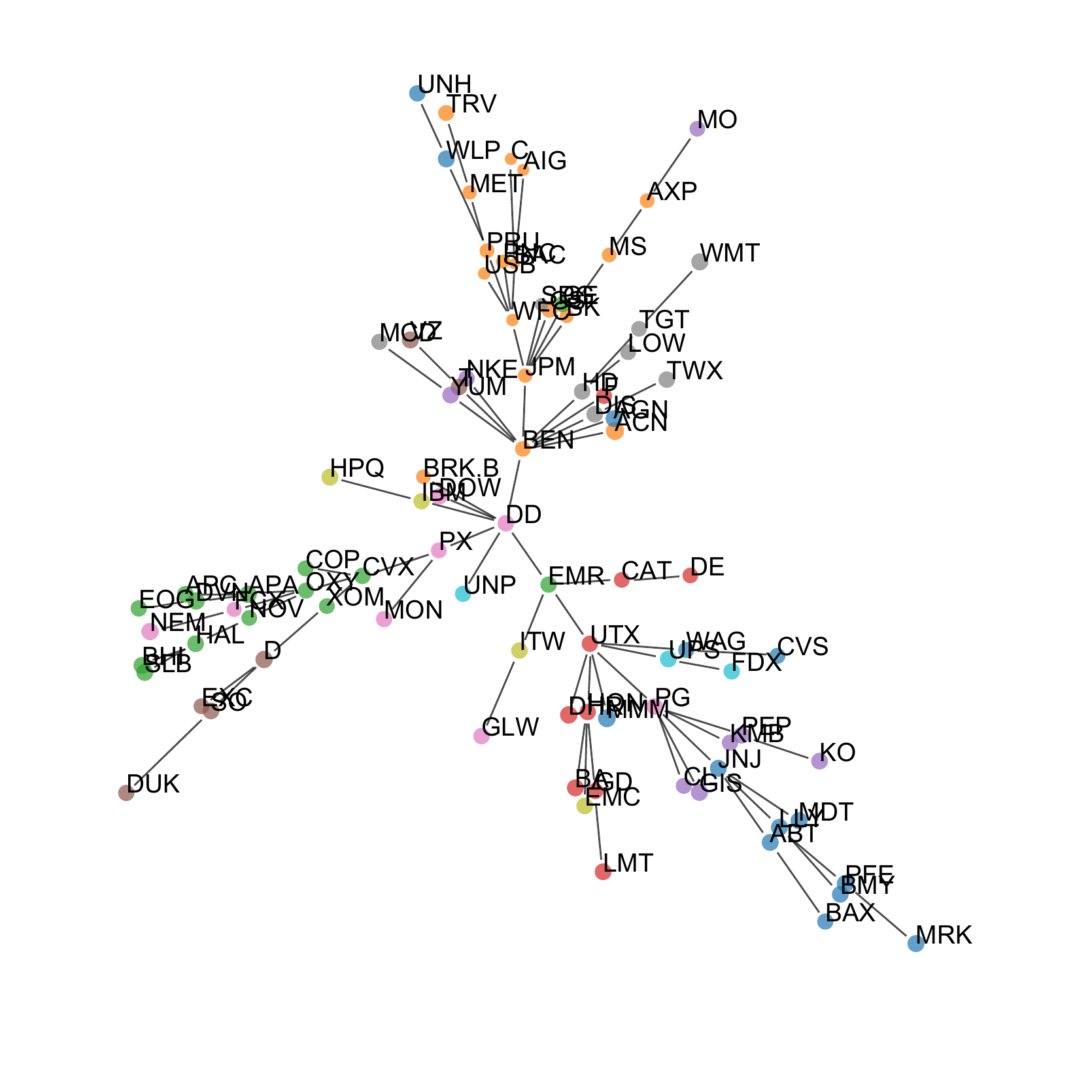}
\caption{{Minimum spanning tree for log price changes (1)}}
\label{fig:network}
\end{figure}

\begin{figure}[tbh]
\centering
\includegraphics[width=0.5\textwidth]{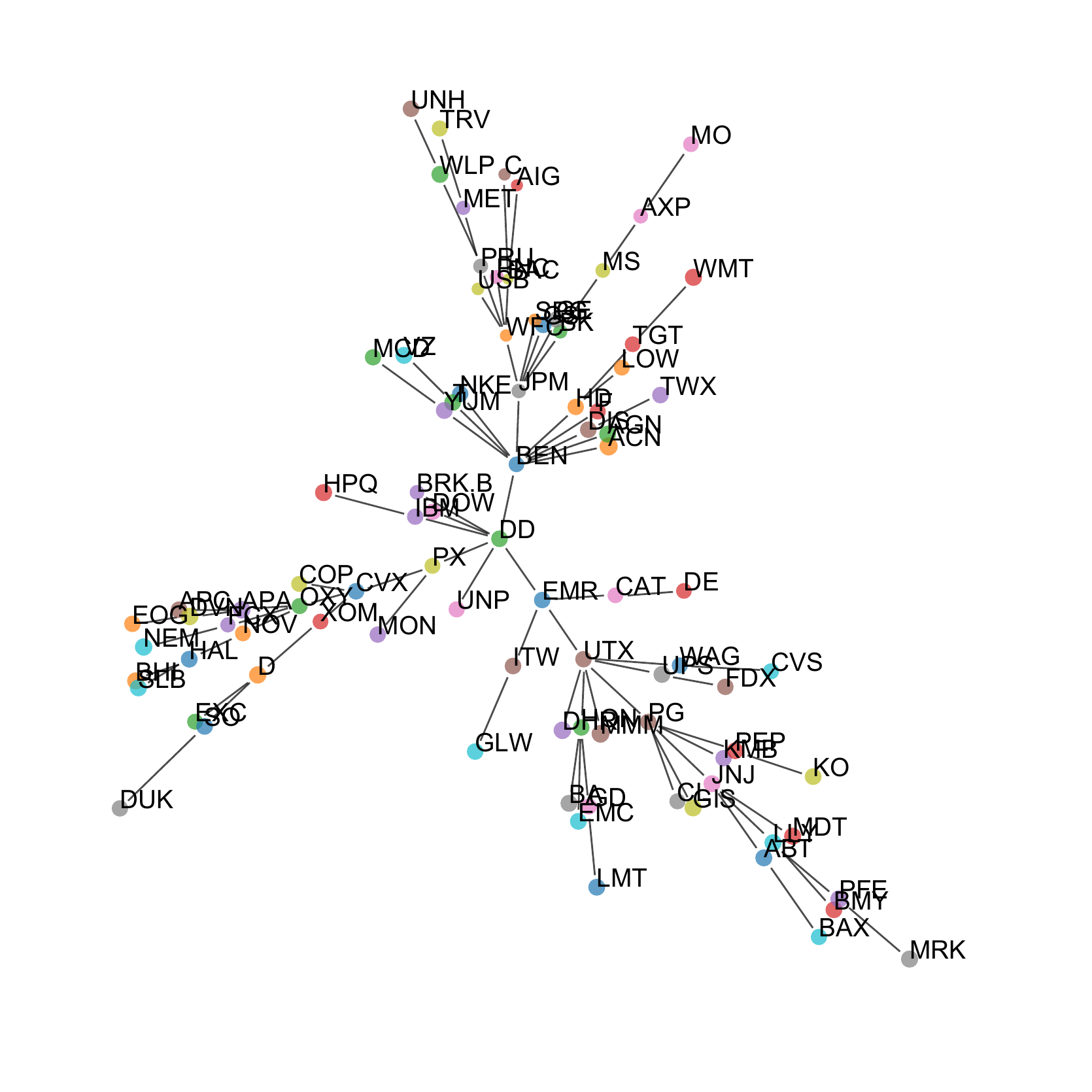}
\caption{{Minimum spanning tree for log price changes (2)}}
\label{fig:networkent}
\end{figure}

\begin{figure}[tbh]
\centering
\includegraphics[width=0.5\textwidth]{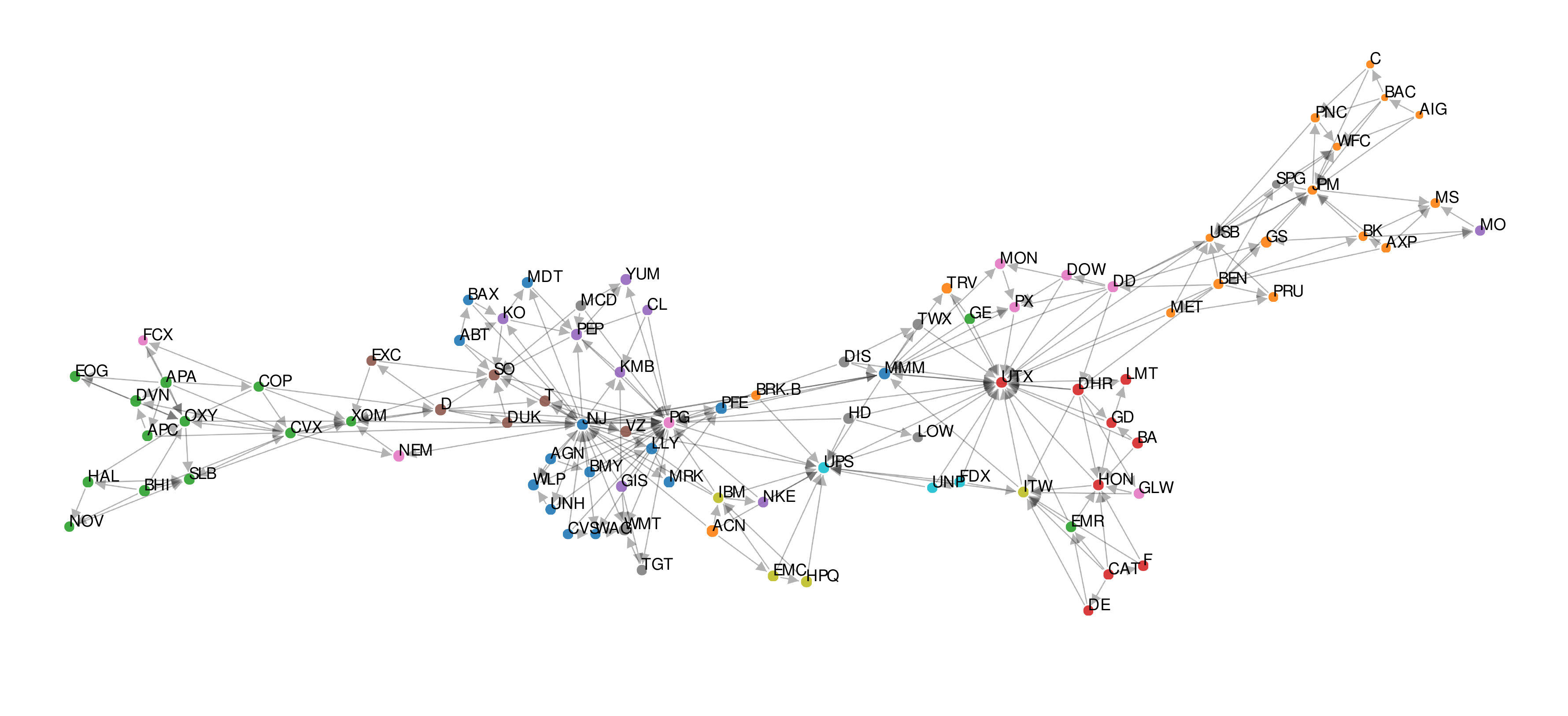}
\caption{{PMFG for log price changes (1)}}
\label{fig:networkP}
\end{figure}

\begin{figure}[tbh]
\centering
\includegraphics[width=0.5\textwidth]{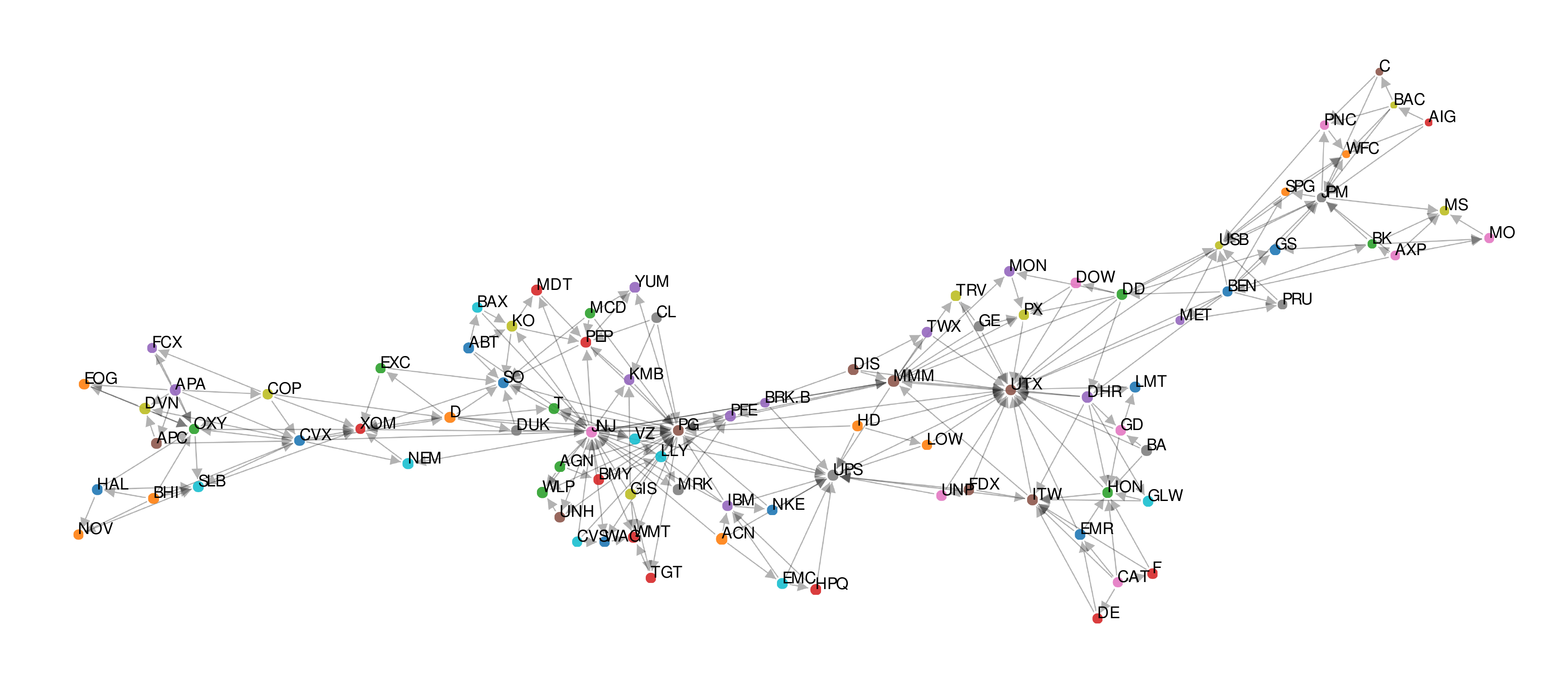}
\caption{{PMFG for log price changes (2)}}
\label{fig:networkentP}
\end{figure}

Minimum spanning trees created for correlations between log price changes are shown on Figures \ref{fig:network} (colours according to sectors) and \ref{fig:networkent} (colours according to entropy rates). Equivalent planar maximally filtered graphs are shown on Figures \ref{fig:networkP} and \ref{fig:networkentP}. This allows us to see that while price changes are dependent on sector in which a company operates the predictability of price changes is not dependent on the sector (entropy rates are not sector dependent). Those figures also show which stocks are key to NYSE 100 index in the sense of being crucial in the propagation of market trends.

\section{Algorithmic trading and profitability}

In this section we will only mention a slight insight into the relationship between predictability of price changes and profitability. Navet \& Chen showed that for price formation processes with higher entropy genetic trading algorithm performed better than lottery trading, while the opposite was true for price formation processes with low entropy \cite{Navet:2008}. Wondering whether the choice of algorithm is meaningful we have performed a calculation using simple mean reversing algorithm (implemented in Python) on the same stocks they have used for the period between 3. Jan 2002 and 31. Dec 2006, which covers most of the period they have used (2001-2006). We compare it not with lottery trading, but with average return on the stocks used, arguing that lottery trading is just randomly sampling the average returns. 

\begin{figure}[tbh]
\centering
\includegraphics[width=0.47\textwidth]{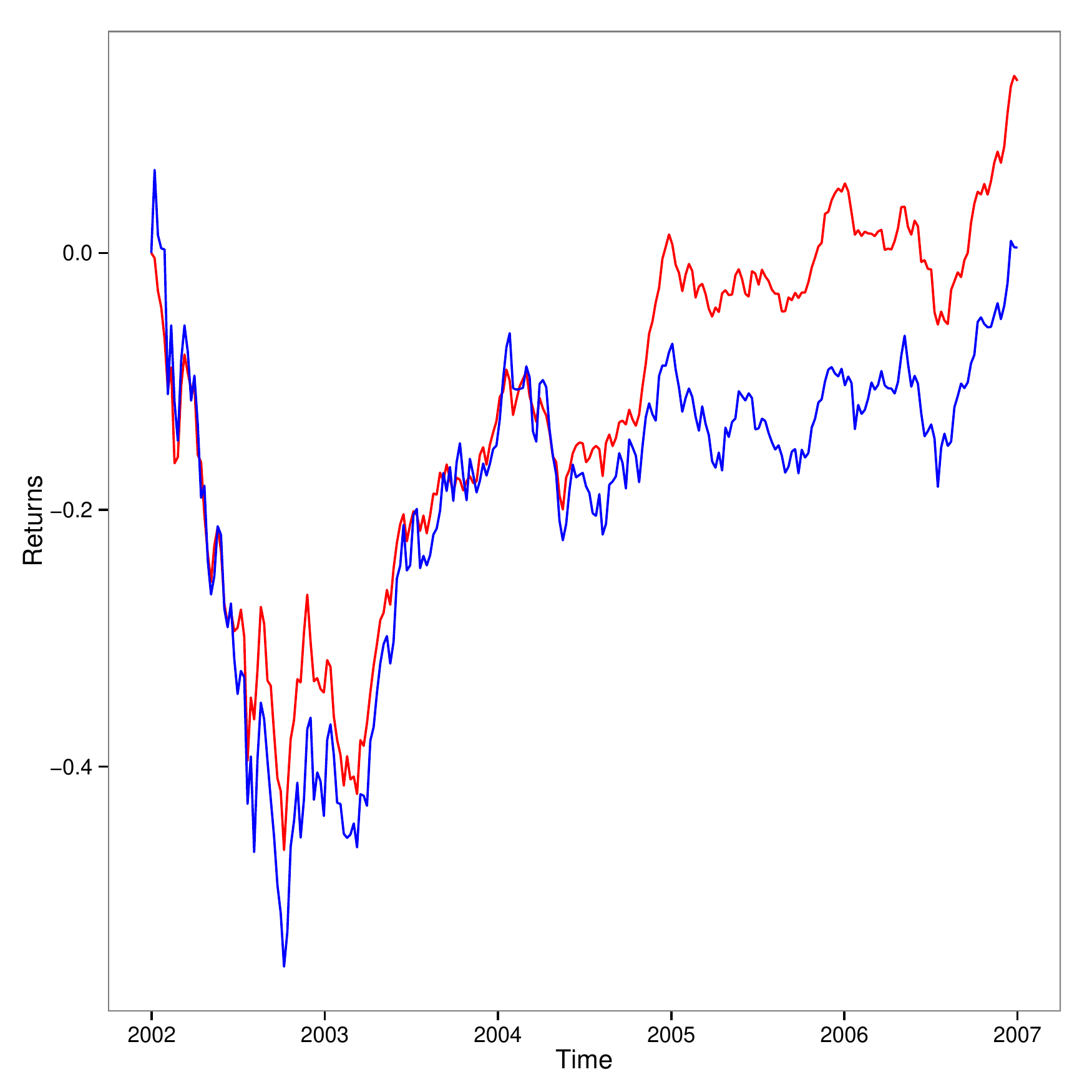}
\caption{{Performance comparison for low entropic stocks}}
\label{fig:LowEntropy}
\end{figure}

\begin{figure}[tbh]
\centering
\includegraphics[width=0.47\textwidth]{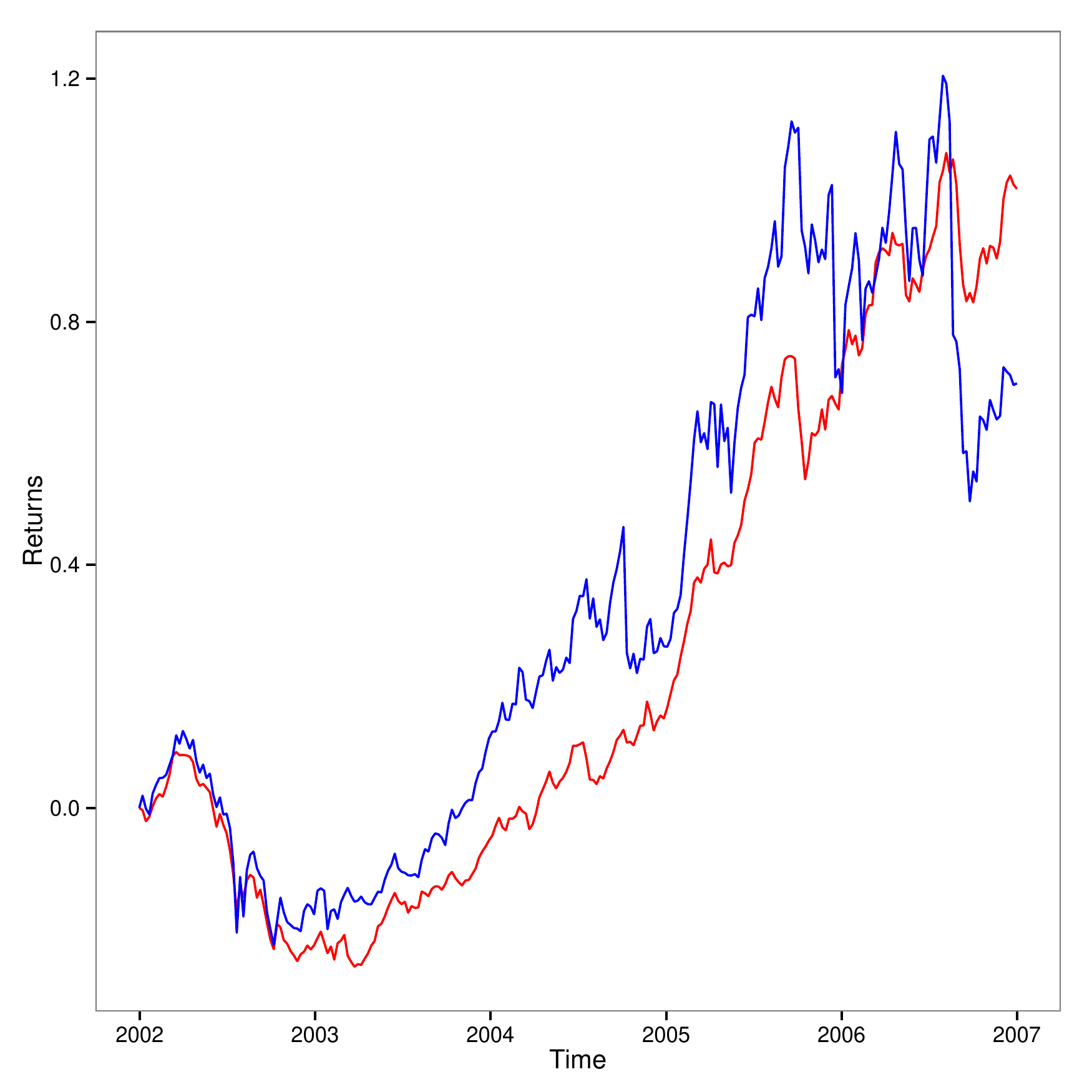}
\caption{{Performance comparison for high entropic stocks}}
\label{fig:HighEntropy}
\end{figure}

The \% returns for mean reversal algorithm (red) and the stocks (blue) for low entropic stocks are shown on Figure \ref{fig:LowEntropy}, and for high entropic on Figure \ref{fig:HighEntropy}. It appears that the algorithm outperforms the market slightly for low entropic price formation, but (even though it has been more profitable in the end) generally underperformed against the market for high entropic processes. Those are clearly different from results obtained by Navet \& Chen, therefore it seems that the relationship between predictability and profitability is not a simple one to capture and this problem will require further exhaustive studies taking into account more trading strategies, which will be rendered possible using intraday price datasets for other markets.

\section{Conclusions}

The results of this study present that the high frequency price changes are slightly but statistically significantly more predictable than daily price changes for NYSE100 stocks. Further studies should be performed to confirm whether this result is robust and appears on other stock markets as well. We also show that predictability of price changes is not dependent on industry (and therefore log returns themselves). Further research should look at whether this is the case on other markets as well. The limited sample for high-frequency data did not allow to study the temporal behaviour of the entropy rates, which should be a subject to further inquiries. The study also shows that profitability, while dependent on predictability, is connected with it in a complex manner, which should be a subject to further research involving various markets and trading strategies. Further research should also look at how different price formation models known in literature are conforming with the market data in terms of the predictability of price changes.

\bibliographystyle{IEEEtran}
\bibliography{IEEEabrv,prace}

\begin{thebibliography}{10}
\providecommand{\url}[1]{#1}
\csname url@samestyle\endcsname
\providecommand{\newblock}{\relax}
\providecommand{\bibinfo}[2]{#2}
\providecommand{\BIBentrySTDinterwordspacing}{\spaceskip=0pt\relax}
\providecommand{\BIBentryALTinterwordstretchfactor}{4}
\providecommand{\BIBentryALTinterwordspacing}{\spaceskip=\fontdimen2\font plus
\BIBentryALTinterwordstretchfactor\fontdimen3\font minus
  \fontdimen4\font\relax}
\providecommand{\BIBforeignlanguage}[2]{{%
\expandafter\ifx\csname l@#1\endcsname\relax
\typeout{** WARNING: IEEEtran.bst: No hyphenation pattern has been}%
\typeout{** loaded for the language `#1'. Using the pattern for}%
\typeout{** the default language instead.}%
\else
\language=\csname l@#1\endcsname
\fi
#2}}
\providecommand{\BIBdecl}{\relax}
\BIBdecl

\bibitem{Lee:2009}
C.-C. Lee and J.~Lee, ``Energy prices, multiple structural breaks, and
  efficient market hypothesis,'' \emph{Applied Energy}, vol.~86, no.~4, pp.
  466--479, 2009.

\bibitem{Yen:2008}
G.~Yen and C.-F. Lee, ``Efficient market hypothesis (emh): Past, present and
  future,'' \emph{Review of Pacific Basin Financial Markets and Policies},
  vol.~11, no.~2, pp. 305--329, 2008.

\bibitem{Rosser:2008}
B.~Rosser, ``Econophysics and economic complexity,'' \emph{Advances in Complex
  Systems}, vol.~11, no.~5, pp. 745--760, 2008.

\bibitem{Mantegna:2000}
R.~N. Mantegna and H.~E. Stanley, \emph{{Introduction to Econophysics:
  Correlations and Complexity in Finance}}.\hskip 1em plus 0.5em minus
  0.4em\relax {Cambridge University Press}, 2000.

\bibitem{Song:2010}
C.~Song, Z.~Qu, N.~Blumm, and A.-L. Barab\'{a}si, ``{Limits of Predictability
  in Human Mobility},'' \emph{Science}, vol. 327, no. 5968, pp. 1018--1021,
  2010.

\bibitem{Navet:2008}
N.~Navet and S.-H. Chen, ``{On Predictability and Profitability: Would GP
  Induced Trading Rules be Sensitive to the Observed Entropy of Time Series?}''
  in \emph{{Natural Computing in Computational Finance}}, ser. Studies in
  Computational Intelligence, T.~Brabazon and M.~O'Neill, Eds.\hskip 1em plus
  0.5em minus 0.4em\relax Springer, 2008, vol. 100.

\bibitem{Youwei:2005}
L.~Youwei and H.~Zue-Zhong, ``{Heterogeneity, Profitability and
  Autocorrelations},'' \emph{{Computing in Economics and Finance}}, no. 244,
  2005.

\bibitem{Kohli:2011}
C.~Kohli and R.~Suri, ``{The price is right? Guidelines for pricing to enhance
  profitability.}'' \emph{Business Horizons}, vol.~54, no.~6, pp. 563--573,
  2011.

\bibitem{Wu:2004}
Y.~Wu, S.~H.~K. Tang, X.~M. Fan, and N.~Groenewold, \emph{The Chinese Stock
  Market: Efficiency, Predictability And Profitability}.\hskip 1em plus 0.5em
  minus 0.4em\relax Cheltenham: Edward Elgar Publishing, 2004.

\bibitem{Rothenstein:2004}
R.~Rothenstein and K.~Pawelzik, ``Limited profit in predictable stock
  markets,'' \emph{Physica A}, vol. 348, pp. 419--427, 2005.

\bibitem{Menkveld:2011}
A.~J. Menkveld, ``High frequency trading and the new-market makers,'' Tinbergen
  Institute, Tinbergen Institute Discussion Papers 11-076/2/DSF21, 2011.

\bibitem{Shannon:1948}
C.~E. Shannon, ``{A Mathematical Theory of Communication},'' \emph{The Bell
  System Technical Journal}, vol.~27, pp. 379--423, 623--656, July, October
  1948.

\bibitem{Cover:1991}
T.~Cover and J.~Thomas, \emph{Elements of Information Theory}.\hskip 1em plus
  0.5em minus 0.4em\relax John Wiley \& Sons, 1991.

\bibitem{Maciejewski:2008}
A.~Maciejewski, M.~Latka, and W.~Jernajczyk, ``{Zastosowanie metody empirycznej
  dekompozycji modalnej i zlozonosci Lempel'a-Ziv'a do analizy EEG chorych na
  schizofrenie},'' \emph{{Medycyna Dydaktyka Wychowanie}}, vol.~XL, no.~1, pp.
  90--94, 2008.

\bibitem{Gao:2006}
Y.~Gao, I.~Kontoyiannis, and E.~Bienenstock, ``From the entropy to the
  statistical structure of spike trains,'' in \emph{IEEE International
  Symposium on Information Theory}, 2006, pp. 645--649.

\bibitem{Farah:1995}
M.~Farah, M.~Noordewier, S.~Savari, L.~Shepp, A.~Wyner, and J.~Ziv, ``{On the
  entropy of DNA: algorithms and measurements based on memory and rapid
  convergence},'' in \emph{SODA'95: Proceedings of the Sixth Annual ACM-SIAM
  Symposium on Discrete Algorithms}.\hskip 1em plus 0.5em minus 0.4em\relax
  Society of Industrial and Applied Mathematics, 1995, pp. 48--57.

\bibitem{Kontoyiannis:1998a}
I.~Kontoyiannis, P.~Algoet, Y.~Suhov, and A.~Wyner, ``{Nonparametric entropy
  estimation for stationary processes and random fields, with applications to
  English text},'' \emph{{IEEE Transactions on Information Theory}}, vol.~44,
  no.~3, pp. 1319--1327, 1998.

\bibitem{Lempel:1977}
A.~Lempel and J.~Ziv, ``{A Universal Algorithm for Sequential Data
  Compression},'' \emph{{IEEE Transactions on Information Theory}}, vol. IT-23,
  no.~3, pp. 337--343, 1977.

\bibitem{Willems:1995}
F.~Willems, Y.~Shtarkov, and T.~Tjalkens, ``{The Context-Tree Weighting Method:
  Basic Properties},'' \emph{{IEEE Transactions on Information Theory}},
  vol.~41, no.~3, pp. 653--664, 1995.

\bibitem{Kennel:2005}
M.~Kennel, J.~Shlens, H.~Abarbanel, and E.~Chichilnisky, ``{Estimating entropy
  rates with Bayesian confidence intervals},'' \emph{{Neural Computation}},
  vol.~17, no.~7, pp. 1531--1576, 2005.

\bibitem{Louchard:1997}
G.~Louchard and W.~Szpankowski, ``{On the average redundancy rate of the
  Lempel-Ziv code},'' \emph{{IEEE Transactions on Information Theory}},
  vol.~43, no.~1, pp. 2--8, 1997.

\bibitem{Leonardi:2010}
F.~Leonardi, ``Some upper bounds for the rate of convergence of penalized
  likelihood context tree estimators,'' \emph{Brazilian Journal of Probability
  and Statistics}, vol.~24, no.~2, pp. 321--336, 2010.

\bibitem{Gao:2008}
Y.~Gao, I.~Kontoyiannis, and E.~Bienenstock, ``Estimating the entropy of binary
  time series: Methodology, some theory and a simulation study.''
  \emph{Entropy}, vol.~10, no.~2, pp. 71--99, 2008.

\bibitem{Doganaksoy:2006}
A.~Doganaksoy and F.~Gologlu, ``{On Lempel-Ziv Complexity of Sequences},''
  \emph{{Lecture Notes in Computer Science}}, vol. 4086, pp. 180--189, 2006.

\bibitem{Kontoyiannis:1998}
I.~Kontoyiannis, ``{Asymptotically Optimal Lossy Lempel-Ziv Coding},'' in
  \emph{IEEE International Symposium on Information Theory}.\hskip 1em plus
  0.5em minus 0.4em\relax Cambridge: MIT, 1998.

\bibitem{Willems:1996}
F.~Willems, Y.~Shtarkov, and T.~Tjalkens, ``Context weighting for general
  finite-context sources,'' \emph{{IEEE Transactions on Information Theory}},
  vol.~42, no.~5, pp. 1514--1520, 1996.

\bibitem{Willems:1998}
F.~Willems, ``The context-tree weighting method: Extensions,'' \emph{{IEEE
  Transactions on Information Theory}}, vol.~44, no.~2, pp. 792--798, 1998.

\bibitem{London:2002}
M.~London, ``The information efficacy of a synapse,'' \emph{Nature
  Neuroscience}, vol.~5, no.~4, pp. 332--340, 2002.

\bibitem{Kennel:2002}
M.~Kennel and A.~Mees, ``Context-tree modeling of observed symbolic dynamics,''
  \emph{Physical Review E}, vol.~66, 2002.

\bibitem{Biggar:2008}
P.~Biggar, N.~Nash, K.~Williams, and D.~Gregg, ``An experimental study of
  sorting and branch prediction,'' \emph{ACM Journal of Experimental
  Algorithmics}, vol.~12, pp. 1--39, 2008.

\bibitem{Zijlstra:2013}
E.~S. Zijlstra, A.~Kalitsov, T.~Zier, and M.~E. Garcia, ``{Squeezed Thermal
  Phonons Precurse Nonthermal Melting of Silicon as a Function of Fluence},''
  \emph{{Physical Review X}}, vol.~3, no.~1, 2013.

\bibitem{Steuer:2001}
R.~Steuer, L.~Molgedey, W.~Ebeling, and M.~Jiménez-Monta\~{n}o, ``{Entropy and
  Optimal Partition for Data Analysis},'' \emph{The European Physical Journal B
  - Condensed Matter and Complex Systems}, vol.~19, no.~2, pp. 265--269, 2001.

\bibitem{Hsieh:1993}
D.~Hsieh, ``Implications of nonlinear dynamics for financial risk management,''
  \emph{Journal of Financial and Quantitative Analysis}, vol.~28, no.~1, pp.
  41--64, 1993.

\bibitem{Brock:1995}
W.~Brock, W.~Dechert, B.~LeBaron, and J.~Scheinkman, ``A test for independence
  based on the correlation dimension,'' \emph{Working papers}, no. 9520, 1995.

\bibitem{Chu:2001}
P.~Chu, ``{Using BDS statistics to detect nonlinearity in time series.}'' in
  \emph{53rd session of the International Statistical Institute}, 2001.

\end{thebibliography}
\end{document}